\documentclass[a4paper,12pt,preprint]{aastex}
\usepackage{natbib}
\usepackage{graphicx}
\usepackage{array}
\usepackage{fancyhdr}
\usepackage{lastpage}
\usepackage[tbtags]{amsmath}
\usepackage{amsthm}
\usepackage{amsfonts}
\usepackage{amssymb}
\usepackage{wasysym} 
\usepackage{arcs} 
\usepackage[d]{esvect} 
\usepackage{color} 
\usepackage{url}
\usepackage[breaklinks=true]{hyperref}
\usepackage{cancel}
\usepackage{comment}
\hypersetup{pdfmenubar=true,pdfstartview={}}
\usepackage{sectsty}
\sectionfont{\normalsize}
\renewcommand{\thesection}{\arabic{section}}
\subsectionfont{\normalsize}
\renewcommand{\thesubsection}{\arabic{section}.\arabic{subsection}}
\usepackage{soul}
\usepackage[left=35mm,right=35mm,top=25mm,bottom=25mm]{geometry}
\def\dg  {^{\circ}}

\newcommand{\AIPS}{{$\cal AIPS$ }}

\newcommand{\com}[0]{{$^{12}$CO}}

\newcommand{\siom}[0]{{$^{28}$SiO}}
\newcommand{\sioi}[0]{{$^{28}$SiO ($J=1{-}0$)}}
\newcommand{\sioii}[0]{{$^{28}$SiO ($J=2{-}1$)}}
\newcommand{\sioiii}[0]{{$^{28}$SiO ($J=3{-}2$)}}
\newcommand{\siov}[0]{{$^{28}$SiO ($J=5{-}4$)}}
\newcommand{\kms}[0]{kms^{-1}}

\title{Dense molecular clumps in the envelope of the yellow hypergiant IRC+10420}
\shorttitle{Dense molecular clumps in IRC+10420}
\shortauthors{Dinh-V-Trung et al.}
\author{Dinh-V-Trung}

\affil{Institute of Physics,
Vietnam Academy of Science and Technology, 10 DaoTan, ThuLe, BaDinh, Hanoi, Vietnam}
\email{dvtrung@iop.vast.ac.vn}
\author{Ka-tat Wong\footnote{Present address: Institut de Radioastronomie
Millim{\'e}trique, 300 rue de la Piscine, 38406
Saint-Martin-d'H{\`e}res, France; wong@iram.fr}}
\affil{Department of Physics, The University of Hong Kong, Pokfulam Road, Hong Kong}
\author{Jeremy Lim}
\affil{Department of Physics, The University of Hong Kong, Pokfulam Road, Hong Kong}
\email{jjlim@hku.hk}
\begin{document}
\begin{abstract}
The circumstellar envelope of the hypergiant star IRC+10420 has been traced as far out in {\sioii} as in {\com} $J = 1$-0 and {\com} $J = 2$-1, in dramatic contrast with the centrally condensed (thermal) SiO- but extended CO-emitting envelopes of giant and supergiant stars.  Here, we present an observation of the circumstellar envelope in {\sioi} that, when combined with the previous observation in {\sioii}, provide more stringent constraints on the density of the SiO-emitting gas than hitherto possible.  
The emission in {\sioi} peaks at a radius of $\sim$2\arcsec\ whereas that in {\sioii} emission peaks at a smaller radius of $\sim$1\arcsec, giving rise to their ring-like appearances.  The ratio in brightness temperature between {\sioii} and {\sioii} decreases from a value well above unity at the innermost measurable radius to about unity at radius of $\sim$2\arcsec, beyond which this ratio remains approximately constant.  Dividing the envelope into three zones as in models for the {\com} $J = 1$-0 and {\com} $J = 2$-1 emission,
we show that the density of the SiO-emitting gas is comparable with that of the CO-emitting gas in the inner zone, but at least an order of magnitude higher by comparison in both the middle and outer zones.  The SiO-emitting gas therefore originates from dense clumps, likely associated with the dust clumps seen in scattered optical light, surrounded by more diffuse CO-emitting interclump gas.
We suggest that SiO molecules are released from dust grains due to shock interactions between the dense SiO-emitting clumps and the diffuse CO-emitting interclump gas.
\end{abstract}

\section{INTRODUCTION}\label{introduction}

Yellow hypergiants are thought to be post-red supergiant stars that are evolving along the blueward loop in the Hertzsprung-Russel (HR) diagram towards higher effective temperatures.   They define the empirical upper limit in luminosity at intermediate stellar effective temperatures (de Jagger 1998, Davidson \& Humphreys 1994).  These stars are surrounded by massive circumstellar envelopes composed of warm dust and molecular gas, providing evidence for increased dynamical instabilities during this poorly understood phase of stellar evolution.  Offering a historical record of their mass loss, studies of the circumstellar envelopes of yellow hypergiants can shed light on the physical processes that drive the mass ejection and affect the evolution of these stars.

One of the best studied yellow hypergiant is IRC+10420.  Based on photometric, spectroscopic, and polarimetric measurements, Humphreys et al. (2002) argue for a distance to this star in the
range 4$-$6\,kpc.  For convenience, we adopt a distance of 5\,kpc for IRC+10420 so that the results reported in this paper can be directly compared with those of both Castro-Carrizo et al. (2007) and Dinh-V-Trung et al. (2009) for its molecular envelope.  In the optical, IRC+10420 was initially classified as F8 I$^{+}_{\rm a}$ in 1973 (Humphreys et al. 1973), but by the 1990s had evolved to spectral type A (Oudmaijer et al. 1996, Klochkova et al. 1997).  Such a rapid change in stellar spectral type is unlikely to reflect an actual change in the stellar surface temperature, but is instead attributed to variations in its optically-thick wind that obscures the stellar surface from view (Humphreys et al. 2002).  Evidence for such a wind comes from optical lines such as the prominent H$\alpha$ line and Ca II triplets, suggesting strong outflow close to the stellar surface (Humphreys et al. 2002). 

Ranking among the brightest infrared sources in the sky, the infrared emission of IRC+10420 constitutes reprocessed stellar radiation by dust in a massive circumstellar envelope ejected over the last $\sim$4000 years.  Observations in the optical with the Hubble Space Telescope (HST) reveal a highly complex and clumpy envelope as seen through scattered light from dust (Humphreys et al. 1997, Tiffany et al. 2010).  Features such as knots and arcs or loops reveal that, on relatively small scales, the mass loss from the star is highly inhomogeneous. 
By combining line-of-sight velocities measured using slit spectroscopy by Humphreys et al. (2002) with proper motions measured from multi-epoch HST images, Tiffany et al. (2010) inferred that the dust clumps with measurable proper motions are moving preferentially on the plane of the sky with velocities of typically $\sim$100$-$$200 {\rm \ km \ s^{-1}}$.  They suggest that the dust is ejected preferentially in an equatorial outflow (i.e., IRC+10420 is viewed closely along its pole), which is also consistent with the measurements of high degrees of linear polarization in the infrared emission from the dust ejecta by Shenoy et al. (2015).
 
The circumstellar envelope of IRC+10420 also is a source of strong molecular line emission. Interferometric observations at high angular resolution in {\com} $J = 1$-0 by \citet{cc07} and {\com} $J = 2$-1 by \citet{dvt09} reveal an envelope that is, on a global scale, approximately spherically symmetric, in stark contrast with the equatorial outflow proposed by Tiffany et al. (2010) for the dust clumps.  Through modelling of the {\com} emission, both \citet{cc07} and \citet{dvt09} found that the molecular envelope of IRC+10420 was produced predominantly during two major mass-loss episodes: a more recent and stronger mass-loss episode lasting for less than 1000\,yr with a strong decrease in mass-loss rate over the last $\sim$300\,yr, together with an older and weaker mass-loss episode lasting roughly 4000\,yr.  The radial ranges of these high and low mass-loss episodes correspond roughly (to the same order of magnitude) with those derived by Humphreys et al. (1997) from mid-infrared images of the dust emission. Recently, using high resolution mid- and far-infrared imaging data Shenoy et al. (2016) found two distinct episodes of mass loss in IRC+10420 with a high average mass loss rate of 2$\times$10$^{-3}$M$_\odot$/yr
until about 2000 yrs ago followed by an order of magnitude decrease in the more recent past, thus confirming the strong variation of
mass loss of the central star. 

In addition to {\com}, molecular lines such as HCN and $^{28}$SiO that have much higher critical densities also have been detected towards IRC+10420 (Quintana-Lacaci et al. 2007).  Interferometric observations at high angular resolutions in {\sioii} by \citet{cc01} reveal a roughly spherically symmetric envelope with a pronounced central depression; i.e., a ring-like shell.  Surprisingly, the {\sioii} emission can be traced out to a radius of $\sim$4\arcsec\ ($\sim$20,000\,AU).  Such a large radial extent in {\sioii} emission is unexpected because SiO molecules ought to condense onto dust grains relatively close to the stellar surface.  For a typical evolved star, condensation of SiO molecules onto dust grains is expected to occur at a temperature of about $900 \text{ K}$ \citep{gail13}, so that essentially all SiO molecules are locked in solid form as silicates beyond a radius of ${\sim}10^{3}$\,AU (e.g. Bujarrabal et al. 1989, Lucas et al. 1992).  The latter is an order of magnitude smaller than the radial extent of the {\sioii} emission from IRC+10420.  Indeed, for the vast majority of evolved stars, SiO emission is observed to arise from just the innermost regions of their envelopes ; e.g., in the archetypical envelope of the AGB star IRC+10216 (Sch{\"o}ier et al. 2006), the size of the SiO-emitting portion of the envelope is many times smaller than that of the CO-emitting envelope.  If the same situation holds for IRC+10420, then, at a distance 5\,kpc, the SiO-emitting region of its envelope should have a radius of just $\sim$0\farcs2, far smaller than the angular resolution employed in the observation by \citet{cc01}

The {\sioii} transition has a critical density of ${\sim}10^{5.5} \text{ cm}^{-3}$.  The latter is significantly higher than the gas density in much of the molecular envelope of IRC+10420 as deduced from modelling of {\com} lines, in which the gas density is inferred to decrease outwards from $\sim$$10^5 \text{ cm}^{-3}$ at an inner radius of $\sim$$2.5 \times 10^{16}$\,cm to $\sim$$10^{2} \text{ cm}^{-3}$ at an outer radius of $4.5 \times 10^{17}$\,cm \citep[e.g.][]{cc07,dvt09,teyssier12}.  If the spatially-extended {\sioii} emission originates from the same gas as is traced in {\com}, then the gas temperature must be elevated to a level not normally found in the envelopes of evolved stars so as to significantly excite the SiO molecule.  Assuming local thermodynamic equilibrium (LTE) and an SiO abundance characteristic of the inner regions of evolved star envelopes where SiO is not locked in dust grains, \citet{cc01} found from radiative transfer modelling that, if excited by the same gas as traced in {\com} $J = 1$-0, a gas temperature of at least $\sim$55\,K is required to produce {\sioii} emission at its observed brightness temperature.

To explain the large spatial extent of the SiO emission around IRC+10420, \citet{cc01} argue that large-scale shocks must be prevalent in the envelope of this star.  Such shocks can be produced when an inner (recently ejected) faster-moving part of the wind ploughs into an outer (ejected in the more distant past) slower-moving part of the wind.  By compressing and heating the gas, shocks can elevate both the gas density and temperature.  Furthermore, shocks can evaporate SiO molecules from silicate dust grains.  The existence of shocks in the envelope of IRC+10420 is supported by the detection of H$_2$O and NH$_3$ lines, which typically arise from hot and dense gas behind shock fronts \citep{teyssier12}.  In addition, the detection of OH masers within $\sim$2\arcsec, corresponding roughly to the angular radius of the observed {\sioii} shell, also suggests a density enhancement in the envelope of IRC+10420 \citep[e.g.][]{bowers84,nedoluha92}.

In this paper, we report interferometric observations of {\sioi} emission from the envelope of IRC+10420.  By comparing our maps with those in {\sioii} obtained by \citet{cc01}, we also derive the ratio in brightness temperature of {\sioii} to {\sioi} as a function of radius through the stellar envelope.  From radiative transfer modelling, we show that the measured line ratios cannot arise from the same gas that emits in {\com}, but requires gas at much higher densities (if not also temperatures, although the latter is not strongly constrained by our measurements).  We compare the inferred physical properties of the envelope with the dust features seen in optical images to determine the likely sources of the SiO emission, and discuss the implications of our results for the nature of the mass loss from IRC+10420.  

The paper is organized as follows.  Readers interested in the observation and data reduction should proceed to $\S\ref{observation}$.  Those interested only in the results can skip ahead to $\S\ref{results}$.  In $\S\ref{model}$, we present our model for the SiO-emitting envelope of IRC+10420.  The implications of the results for our understanding of the molecular gas and dust envelope of IRC+10420 is discussed in $\S\ref{discussion}$.  A concise summary of our work can be found in $\S\ref{summary}$.

\section{DATA}\label{observation}
\subsection{Observations}
We observed IRC+10420 with the Karl G. Jansky Very Large Array (VLA) on 19 April 2010 in its most compact (D) configuration. The observation spanned ${\sim}1$ hour, with ${\sim}25$ minutes on the source.  The correlator was configured to span the {\sioi} line (rest frequency of $43.423853$ GHz) over a bandwidth of 32 MHz ($\sim$220 kms$^{-1}$), sufficient to span the entire linewidth of $\sim$$80 {\rm \ km \ s^{-1}}$ observed for the {\sioi} emission towards IRC+10420.  To correct for gain errors introduced by atmospheric fluctuations, which can be quite problematic at high observing frequencies at the VLA, the telescope was switched alternately between IRC+10420 and a nearby secondary calibrator, J1924+1540 (lying $4.36\dg$ away).  In each 5-min cycle, we integrated for ${\sim}160$ s on IRC+10420 and ${\sim}100$ s on J1924+1540.  Another strong quasar, J1642+3948 (also known, and referred to hereafter, as 3C\,345), was observed at the start of the observation for bandpass calibration. 

\subsection{Absolute Flux Calibration}\label{flux calibration}
Due to restrictions imposed by dynamic scheduling, the selected flux calibrator, J0137+3309 (3C\,48), was not observed during the session.  We therefore searched the data archive for other observations of the bandpass calibrator used in our observation, 3C\,345, taken in the same telescope configuration.  Provided that these observations were accompanied by observations of a standard absolute flux calibrator, we could then use 3C\,345 as an absolute flux calibrator (in addition to it serving as a bandpass calibrator).  In Table \ref{table1}, we list the three such observations that we found taken close in time to (within 9 days of) our observation.  One observation (labelled observation 1) was taken before our observation (observation 2), and another two (observations 3 and 4) after our observation.  The data quality in observation 3 was poor, and therefore excluded from further consideration.  We reduced the data taken in observations 1 and 4 to determine the flux density of 3C\,345 in those observations; the data reduction scheme used is the same as that used for reducing our data as described below ($\S\ref{data reduction}$).  In both these observations, the flux calibrator observed was J1331+3030 (also known, and hereafter referred to, as 3C\,286).  Monitoring of 3C\,286 by the observatory has shown that its flux density has remained constant (within ${\sim}2\%$) over the past few decades in all the VLA frequency bands. Note that at the time of all the aforementioned observations, the switched power (SY) table containing information of system gain variations was not available.  Without the SY table, the uncertainty in the flux bootstrapping is roughly $10\%$. 

As listed in Table \ref{table1}, the flux density of 3C\,345 increased by about 38\% (much larger than the uncertainty in flux bootstrapping) between observations 1 and 4, separated by just 16 days.  The  corresponding average daily increase in flux density is ${\sim}0.14$ Jy/day.  Historically, 3C\,345 has been found to undergo periodic major flaring events at radio wavelengths every 8 to 10 years \citep{3c345period}. \citet{3c345flaring} and \citet{3c345gammaray} have shown that 3C\,345 was undergoing a major flaring episode starting from 2008 and lasting until at least mid-2010.  Observations of 3C\,345 with the Very Long Baseline Array (VLBA) in the months around April 2010, as tabulated in Table \ref{table2}, show that 3C\,345 was varying especially rapidly in flux density during April 2010.  From a linear interpolation between the observations, we find that the average daily rate in flux density variations as measured with the VLBA was sometimes even higher than that measured with the VLA as described above.  \citet{3c345gammaray} attribute the flux density variations observed with the VLBA to newly identified ``moving emission features'' in the radio-emitting jet of 3C\,345.  Notice that the flux density of 3C\,345 as measured with the VLBA appeared to decrease in the interval from 6 April to 14 June 2010, whereas that measured with the VLA appeared to increase from 10 April to 26 April 2010.  This difference could reflect the vastly different angular scales of the emitting regions probed by the VLA and VLBA.  On the other hand, the flux density of 3C\,345 could have varied over shorter timescales than can be tracked with the available observations.   If we assume that the flux density of 3C\,345 varied linearly in time between observation 1 and 4, then from a linear interpolation we obtain a flux density of 7.38\,Jy for this source during our observation. Assuming that the flux density of 3C345 did not vary beyond the range used in the interpolation, the uncertainty in the estimate flux density is <=20 

\subsection{Data Reduction}\label{data reduction}
We reduced our VLA data using the NRAO Astronomical Image Processing System, \AIPS (version \texttt{31DEC11}).  First, we corrected for changes in the atmospheric opacity as a function of elevation, for which the default seasonal atmospheric opacity model was used to estimate the zenith atmospheric opacity at 43 GHz.  We also corrected for the changes in the collecting area of the antennas with elevation due to gravity-induced deformation, as well as errors in antenna positions based on measurements made by the observatory.  We then corrected for errors in the geometrical delay between antennas by fitting a slope to the visibility phase measured for 3C\,345 (the bandpass calibrator) across the passband of each antenna.

To solve for the bandpass and complex gain of each antenna, we adopted an iterative approach where we first solved for the complex gain at short time intervals over a limited bandwidth, and applied the complex gain solutions to solve for the bandpass.  We then solved for the complex gain again but now over the entire bandwidth, and applied the complex gain to refine the solution for the bandpass.  This process was repeated until successive solutions showed little change in either the bandpass or complex gain.  The final bandpass solution was applied to both the secondary calibrator and source.  After that, we solved for the complex gain of the secondary calibrator, and derived its flux density with reference to 3C\,345 (see $\S\ref{flux calibration}$).  Finally, we applied the complex gain solutions derived for the secondary calibrator to the target source by interpolating between scans of the secondary calibrator.

Averaging the visibilities over frequency to a velocity resolution of $4.86$ kms$^{-1}$, we first applied a fourier transform to the calibrated visibilities to form DIRTY maps.  We then deconvolved the point spread function of the telescope from the DIRTY maps using CLEAN algorithm \citep{clarkclean} to obtain the final CLEAN maps.   We made maps using ROBUST weighting ($\texttt{ROBUST} = 0$) of the visibilities to achieve a good compromise between angular resolution and the root-mean-square (rms) noise fluctuations.  The synthesized beam thus obtained is $1''.78 \times 1''.36$ with a position angle (PA) of $-44.51\dg$.  The rms noise level, estimated from the DIRTY maps in the line-free channels, is ${\sim}5.9$\,mJy, corresponding to ${\sim}1.1$\,K; i.e., a conversion factor of 185.63 K/Jy. 

Previously published maps of the {\sioii} emission from IRC+10420 made with the Plateau de Bure Interferometer (PdBI) \citep{cc01} were kindly provided to us by A. Castro-Carrizo.  To be able to compare these maps with ours so as to derive the line ratio in {\sioii} to {\sioi}, we first regridded the {\sioii} maps to the same velocity resolution as our {\sioi} maps.  In addition, we convolved our {\sioi} maps to the slightly larger synthesized beam of the {\sioii} maps of $2''.53 \times 1''.38$ and a position angle of $\text{PA} = 25.98\dg$.  From these maps, we derived the line ratio of {\sioii} to {\sioi}, hereafter referred to as ${\text {\sioii}}/{\text {\sioi}}$, as a function of radius averaged over circular annulli.

\section{RESULTS}\label{results}
Figure\,1 shows our channel maps of the {\sioi} emission from IRC+10420.  For comparison, Figure\,2 shows channel maps of the {\sioii} emission made by \citet{cc01} after being regridded to the same velocity resolution as our {\sioi} maps. For reference, the cross marks the location of the central star as detected in the millimeter continuum by Dinh-V-Trung et al. (2009).  In both maps, the emission has the characteristic signature of an expanding spherical envelope, as seen previously also in {\com}: the emission exhibits the largest spatial extent in velocity channels around the systemic velocity of 74 kms$^{-1}$, and contracts in size towards the center of the envelope (position of the star) at progressively higher blueshifted and redshifted velocities.  The {\sioi} emission spans a velocity range between 38 kms$^{-1}$ and 120 kms$^{-1}$, corresponding to an expansion velocity of about 41 kms$^{-1}$.  The inferred expansion velocity in {\sioi} is similar to that found previously in {\sioii} \citep{cc01} as well as in {\com} \citep{cc07,dvt09}.  The diameter of the {\sioi} emission as shown in Figure 1 is about 8\arcsec, comparable to that mapped in {\sioii} by \citet{cc01} and in {\com} $J = 2$-1 by \citet{dvt09}.  The combined single-dish and interferometer map in {\com} $J = 1$-0 by \citet{cc07} traces the envelope somewhat further out to a diameter of about 12\arcsec.

The higher angular resolution maps in {\sioi} reveal more internal structure than was previously seen in {\sioii}.  At and near the systemic velocity, the {\sioi} emission exhibits a clumpy ring-like structure centered approximately at the position of the central star.  The emission from the south-east part of the shell is clearly weaker than that from the north-west part, indicating a departure from spherical symmetry at least in {\sioi}.  A ring-like structure also is seen in the {\sioii} maps of \citet{cc01} at and near to the systemic velocity.  As we will show below, however, the {\sioii} ring is significantly smaller than the {\sioi} ring.  The asymmetry in brightness around the ring is not as conspicuous in the {\sioii} as in the {\sioi} maps, although this difference may be related to the different angular resolutions of the two maps.

At extreme blueshifted velocities (43.3--$53.6 {\rm \ km \ s^{-1}}$), the centroid of the {\sioi} emission is located to the south-west of the central star.  By comparison, at extreme redshifted velocities (105.4--$110.6 {\rm \ km \ s^{-1}}$), the centroid of the {\sioi} emission is located to the north-east of the central star.  In the maps at a lower angular resolution in {\sioii}, the same shift also is apparent at extreme blueshifted velocities although not at extreme redshifted velocities (where the centroid is coincident with the central star).  In Figure\,3, we show a position-velocity diagram in {\sioi} for a cut along the north-east to south-west direction through the central star at a position angle of PA = 70$^\circ$.  A velocity gradient is apparent in this direction as judged by the centroid of the emission at extreme blueshifted and redshifted velocities as described above.  A similar velocity gradient was previously seen also in {\com} $J = 1$-0 by \citet{cc07} and {\com} $J = 2$-1 by \citet{dvt09}, indicating that the SiO-emitting and CO-emitting gas share the same global spatial-kinematic structure. 

In Figure\,4, we show the radially-averaged brightness temperature profiles in {\sioi} and {\sioii} from maps regridded to the same velocity resolution and convolved to the same angular resolution in both lines ($\S\ref{data reduction}$).  These profiles are obtained by azimuthally averaging the line emission over successive annuli of 0\farcs7, which corresponds approximately to Nyquist sampling.  The error bars indicate only random uncertainties, and do not include systematic uncertainties in the flux calibration of either the {\sioi} ot {\sioii} lines.
The brightness temperature in {\sioi} peaks at a radius of $\sim$2\arcsec, whereas that in {\sioii} peaks at a significantly smaller radius of $\sim$1\arcsec.  Throughout the inner $\sim$2\arcsec, the brightness temperature in {\sioii} is significantly higher than that in {\sioi}.  Figure\,5 shows the ratio in brightness temperature between {\sioii} and {\sioi} as a function of radius, computed from the brightness temperature profiles of these two lines as shown in Figure\,4.  As can be seen, the line ratio {\sioii}/{\sioi} reaches its highest value of 2.2 $\pm$0.2 at the innermost annuli.  On the other hand, at larger radii, both lines have comparable brightness temperatures, such that {\sioii}/{\sioi} is practically unity at radii beyond $\sim$2\arcsec.  The different line ratios over different radial ranges indicate a difference in the excitation of the {\sioi} and {\sioii} lines at the inner compared to the outer regions of the envelope.  

Systemic errors in the flux calibration, as described in $\S\ref{flux calibration}$ for our observation in {\sioi}, may affect the measured brightness temperatures of both the {\sioi} and {\sioii} lines and hence their measured ratio in brightness temperatures.  Such errors, however, do not change the observed trends in the brightness temperatures or the ratio in brightness temperatures of both these lines as a function of radius as described above.  In our model for reproducing the azimuthally-averaged brightness temperatures and ratio in brightness temperatures of these lines as a function of radius as described next, systematic errors in flux calibration may lead to a quantitative change in the physical parameters derived, but not a qualitative understanding of the differences in physical properties between the SiO-emitting and CO-emitting envelope.

\section{MODEL}\label{model}
As we show below, by combining our measurements in {\sioi} with those in {\sioii} by \citet{cc01} as described in $\S\ref{results}$, we are able to better constrain the density of the molecular hydrogen (H$_2$) gas responsible for exciting the observed SiO lines than was possible from the {\sioii} line alone.

\subsection{Qualitative Assessment}\label{RADEX}
As a preliminary and qualitative assessment of how {\sioii}/{\sioi} varies with physical conditions within the envelope of IRC+10420, we make use of the simple radiative transfer code \texttt{RADEX} \citep{radex}.  \texttt{RADEX} adopts the large velocity gradient (LVG) or Sobolev approximation \citep{sobolev60}, whereby photons emitted by molecules at a given location will not be absorbed by those elsewhere along the line of sight due to a Doppler shift caused by a velocity change \citep{kwok07}.  As a consequence, radiative transfer can be treated as a local problem, vastly simplifying the computation of radiative transfer through a parcel of gas.  Assuming a constant H$_2$ gas density and a constant SiO abundance, Figure 6 shows how {\sioii}/{\sioi} is predicted by \texttt{RADEX} to vary with H$_2$ gas density, $n_{{\text{H}}_2}$ (cm$^{-3})$, and SiO column density, $N_{\text{SiO}}$ (cm$^{-2}$), at a gas temperature of 100\,K, which is within the range of temperatures previously estimated for the inner envelope of IRC+10420.  Two general trends are apparent in this figure, such that {\sioii}/{\sioi} increases as: (1) $n_{{\text{H}}_2}$ increases; and (2) $N_{\text{SiO}}$ decreases.  As $n_{{\text{H}}_2}$ increases, collisional excitation by H$_2$ molecules increasingly populate the $J = 2$ level at the expense of the $J = 1$ level, leading to an increase in {\sioii}/{\sioi}.  As $N_{\text{SiO}}$ increases, the optical depth increases, and the excitation temperature of {\siom} more closely approaches the gas kinetic temperature as radiative trapping contributes increasingly to the excitation.  As a consequence, {\sioii}/{\sioi} approaches a value of unity as $N_{\text{SiO}}$ increases.

Attaining {\sioii}/{\sioi} in the range $\sim$1--2, as is observed, requires $n_{{\text{H}}_2} \gtrsim 10^{4-5} \rm \ cm^{-3}$.  Except in the inner region of the envelope, such H$_2$ gas densities are significantly higher than that required to produce both the {\com} $J = 1$-0 and {\com} $J = 2$-1 \citep{dvt09} lines according to the models proposed by both \citet{cc07} and \citet{dvt09}.  
Note that \texttt{RADEX} predicts {\sioii}/{\sioi} to be quite insensitive to the gas temperature, and hence our measurements do not provide strong constraints on temperature of the SiO-emitting gas. 

\subsection{Radiative transfer code}\label{Code}
To better constrain physical conditions in the $^{28}$SiO-emitting envelope of IRC+10420, we used the one-dimensional radiative transfer code developed by \citet{dvt00} that was previously used to model the CO-emitting envelope of this star.  Instead of using the LVG approximation as in \texttt{RADEX}, we directly solved the coupled problem of radiative transfer and SiO level population throughout the envelope as a whole.  To represent the stellar envelope, we used a grid of 90 radial mesh points.  The intensity of the radiation field is calculated for each grid point through solving the radiative transfer equation, and the level population of SiO molecules in each grid point computed by solving the statistical equations.  This process was repeated iteratively until convergence was reached, using the accelerated $\Lambda$-iteration method to speed up the calculations.  Once the level population at each grid point was determined, the resultant intensity of a given $^{28}$SiO line was found by integrating the radiative transfer equation along a large number of rays through the envelope.

To simulate the observations, we convolved the predicted intensity profile in the sky plane with a circular Gaussian beam.  The diameter of the convolving beam is $1''.87$, which is equal to the geometric mean of the major and minor axes of the synthesized beam ($2''.53 \times 1''.38$).  Collisional cross-sections between $^{28}$SiO and H$_2$ were taken from \citet{lamda}, who scaled (by 1.38) and interpolated the collisional cross-sections of {\siom} with He as computed by \citet{siocolrate}.  We included in our calculations all rotational transition levels of {\siom} up to $J=20$ in the $v = 0$ and $v = 1$ vibrational states.  Collisional excitation to the $v = 1$ vibrational state requires unrealistically high densities and temperatures, and was therefore ignored in our calculations.   On the other hand, strong infrared emission by hot dust located close to the star can contribute significantly to exciting $^{28}$SiO; in particular, $^{28}$SiO molecules in any rotational level $J$ in the $v = 0$ vibrational state may be excited to levels $J \pm 1$ in the $v = 1$ vibrational state by absorbing infrared photons at 8 $\mu$m emitted by the hot inner dust shell.  These molecules are then de-excited very rapidly through spontaneous radiative transitions to rotational levels $J\pm 2$ in the $v = 0$ vibrational state.  For the sake of simplicity, we assume that all the infrared photons at 8 $\mu$m come from an optically thick shell with a temperature of $T = 400$\,K located near the stellar surface, as in \citep{dvt09}.  The estimated temperature of the dust shell is based on fitting the spectral energy distribution from the near-infrared to millimeter wavelengths.  From the observed flux density at 8 $\mu$m from IRC+10420, we estimate the radius of this dust shell to be $\sim$$8 \times 10^{15}$\,cm ($\sim$500\,AU), which is a factor of three smaller than the inner radius of the SiO envelope.

\subsection{Model assumptions}\label{model assumptions}
Despite modelling efforts and recent constraints from observations with the Herschel telescope, the temperature profile in the gaseous envelope of IRC+10420 remains highly uncertain.  In their model of the CO-emitting envelope, \citet{cc07} adopted an analytical profile for the temperature of the stellar envelope.  This profile leads to a relatively high temperature at the inner region of the envelope, reaching about 1200\,K at its innermost boundary (within which the density drops off dramatically due to the drastic decrease in stellar mass-loss rate over the last $\sim$300\,yr).  In contrast, by considering the energy balance between heating and cooling processes, \citet{dvt09} derived in a self-consistent manner the gas temperature throughout the envelope of IRC+10420.  The temperature derived by \citet{dvt09} for the inner region of the envelope is significantly lower than that adopted by \citet{cc07}.  Relying on matching single-dish observations of $^{12}$CO from $J = 1-0$ to $J = 6-5$ and $^{13}$CO from $J = 1-0$ to $J = 2-1$, however, the temperature derived by \citet{dvt09} for the inner region of the envelope is not well constrained (the observed CO transitions arise primarily from the outer, cooler region of the envelope).  More recently, from observations in the $J = 6-5$, $J = 10-9$, and $J = 16-15$ transitions of $^{12}$CO and $^{13}$CO with the Herschel telescope, \citet{teyssier12} inferred a gas temperature at the innermost layers of the envelope that is higher than that derived by \citet{dvt09} and closer to that adopted by \citet{cc07}.  

As mentioned in $\S\ref{RADEX}$, our simple calculations using \texttt{RADEX} indicates that the line ratio {\sioii}/{\sioi} does not depend sensitively on the gas temperature. Thus, for the sake of simplicity, we adopted the temperature profile proposed by \citet{teyssier12} for the inner region and the analytical profile adopted by \citet{cc07} for the outer region of the envelope, parameterised as:

\begin{eqnarray}
\biggl( \frac{T_{\text{kin}}}{\text{K}} \biggr)_{\text{inner}}(r) &= 170 \cdot \biggl( \frac{r}{10^{17}\text{ cm}} \biggr)^{-1.2} \ \ \  {\rm for} \ \ \  0.25 \le \biggl( \frac{r}{10^{17}\text{ cm}} \biggr) \le 1.24 \\
\biggl( \frac{T_{\text{kin}}}{\text{K}} \biggr)_{\text{outer}}(r) &= 100 \cdot \biggl( \frac{r}{10^{17}\text{ cm}} \biggr)^{-0.8}  \ \ \  {\rm for} \ \ \  1.24 < \biggl( \frac{r}{10^{17}\text{ cm}} \biggr) \le 5.20 
\end{eqnarray} 
\noindent  As emphasized by \citet{dvt09}, the gas temperature in the envelope of IRC+10420 is relatively high compared with that in the envelopes of AGB stars.  They suggest that, due to the high stellar luminosity and hence radiation pressure, dust grains are driven outwards at a relatively high velocity.  Collisions between gas molecules and the dust grains then heat up the gas in the envelope to a relatively high temperature.

To account for the smooth line profile of the envelope despite its unusually large expansion velocity of $\sim$$38 \text{ km s}^{-1}$, we adopted a local turbulent velocity of $3 \text{ km s}^{-1}$ as in \citet{dvt09}.  By comparison, the typical expansion velocity in the envelope of AGB stars is $\sim$$10 \text{ km s}^{-1}$, where the turbulent velocity is $\sim$$1 \text{ km s}^{-1}$. We find that our model results do not change significantly for reasonable values of the local turbulent velocity.

\subsection{Model results}\label{model results}

The radial dependence in H$_2$ gas density and SiO abundance are free parameters that we varied until a satisfactory match was found to both the measured radial dependence in {\sioii}/{\sioi} and brightness temperatures of both {\sioi} and {\sioii}.  We emphasize at this point that our model strives to match the azimuthally averaged measurements in {\sioi} and {\sioii}, and makes no attempt to reproduce the clumpiness and departure from circular symmetry seen in both lines.  Our model also predicts the intensity of the SiO $J = 3-2$ and $J=5-4$ lines as observed by \citet{ql07}, helping us assess the viability of the selected model parameters.

\subsubsection{Different physical parameters for SiO- and CO-emitting gas}
Using the H$_2$-gas densities derived by either \citet{cc07} or \citet{dvt09} for the CO-emitting envelope and assuming a reasonable SiO abundance (ratio of SiO to H$_2$ molecules) in the range between 10$^{-7}$ to a few 10$^{-6}$, we found that the predicted brightness temperatures of both {\sioi} and {\sioii} are much lower than are observed except in the innermost region (radius $< 2\arcsec$) where the H$_2$-gas density is close to the critical density of both {\sioi} and {\sioii}.  This result, assuming a gas temperature for the envelope as described in $\S\ref{model assumptions}$, is simple to understand.  Although radiative excitation by IR photons emitted from hot dust near the star is significant, collisions with H$_2$ molecules nevertheless dominates the excitation of both {\sioi} and {\sioii}.  Because the critical densities of both these lines are much higher than that of either CO (J = 1-0) or CO (J = 2-1) on which the models of both \citet{cc07} and \citet{dvt09} are based, {\sioi} and {\sioii} are only weakly excited at the densities inferred for the CO-emitting envelope.  In this situation, both these lines are very optically thin, yielding a line ratio well above unity along all lines of sight through the envelope.  Such a high line ratio is in clear contradiction with that measured along lines of sight at angular radii $\gtrsim 2\arcsec$ ($\gtrsim 10^4 {\rm \ AU}$) from the star.  Now, increasing the gas temperature can potentially boost the brightness temperatures of both lines to a level that matches the values observed, as in the model of \citet{cc01} for the {\sioii} line.  The dramatic increase in gas temperature required, however, would result in much brighter CO lines than are observed \citep[for the gas temperature required to produce the observed CO lines, see the calculations of][]{dvt09}.  

Our central conclusion therefore is that, except at the innermost regions ($< 2\arcsec$), the same gas component cannot be responsible for both the SiO and CO emission from the envelope of IRC+10420.  Instead, the SiO-emitting gas must have a significantly higher density or temperature, or both, than the CO-emitting gas.  

\subsubsection{Physical parameters of SiO-emitting gas}
In the following, we explore the situation in which the SiO-emitting gas has a temperature that is comparable with that of the CO-emitting gas; i.e., a radial dependence in temperature as described in $\S\ref{model assumptions}$.  In $\S\ref{discussion}$, we explain the physical motivation for making this assumption.  Irrespective of the exact gas temperature, as demonstrated in $\S\ref{RADEX}$, the H$_2$ gas density required for producing the observed {\sioi} and {\sioii} emission must be close to or exceed the critical densities of these lines throughout the envelope.  As a matter of convenience for inputting the model parameters as well as to provide an instructive comparison with the CO-emitting envelope, we express the required H$_2$ gas density in the SiO-emitting envelope in terms of a corresponding mass-loss rate (which can change as a function of time) from the central star.  Implicit in this expression is that the mass-loss rate is computed for a unity filling factor for the SiO-emitting gas.  The required H$_2$ gas density for producing the CO-emitting envelope is also expressed as a corresponding mass-loss rate, again assuming implicitly a unity filling factor for the CO-emitting gas, in the models of \cite{cc07} and \cite{dvt09}.  Of course, a star cannot have two different mass-loss rates at the same time, so that any difference in this rate simply indicates different H$_2$ gas densities for the SiO- and CO-emitting gas at a given radius.  The total mass-loss rate of the star is the sum of the individual contributions from the SiO- and CO-emitting parts of the envelope, and depend on their individual filling factors.

As mentioned in $\S\ref{introduction}$, the CO-emitting gas originates from two shells, an inner detached shell and an outer shell, separated by a gap.  Similarly, we divide the SiO-emitting envelope into three zones, an inner zone coinciding with the inner detached shell, an outer zone coinciding with the outer detached shell, and a middle zone corresponding to the gap between these two shells, chosen to match specifically the model for the CO-emitting envelope proposed by \citet{cc07} (given that we adopt their temperature profiles beyond the inner zone).
In the inner zone, spanning a radius of $\sim$0\farcs3 to $\sim$1\farcs5, the {\sioii} emission peaks in brightness temperature.  The {\sioi} emission, however, peaks in brightness temperature further out, in the middle zone.  Thus, in the inner zone, {\sioii}/{\sioi} is relatively high and consistently above unity, implying that both lines are optically thin.  As a consequence, the {\siom} column density is required to be relatively low, and hence also the {\siom} abundance. 
In the middle zone spanning $\sim$1\farcs5 to $\sim$2\farcs5, {\sioii}/{\sioi} is close to but consistently above unity, indicating that both lines have a greater optical depth than the inner region.  As a consequence, the {\siom} column density and hence also abundance must be higher than their corresponding values in the inner zone.  
In the outer zone, ranging from $\sim$2\farcs5 to the outer edge of the envelope at $\sim$6\farcs0, the brightness temperatures of both {\sioi} and {\sioii} decrease smoothly with increasing radius, and have values much lower than those in the middle or inner zone.  Throughout this zone, {\sioii}/{\sioi} has a value consistent with unity, indicating that both lines are somewhat more optically thick than in the middle zone.  Given the relatively large depth of the outer zone but the much lower brightness temperatures in both {\sioi} and {\sioii} by comparison with the middle zone, the {\siom} abundance in the outer zone must be much lower than that in the middle zone. 

Assuming a constant mass-loss rate in each of the three zones of the SiO-emitting envelope just like for the CO-emitting envelope, we found a satisfactory match between our model and the results in both {\sioi} and {\sioii} if the SiO-emitting envelope has the following mass-loss rate in the three zones:   

\begin{equation}
\dot{M} (M_\odot/yr) = \left\{
\begin{array}{lr}
1.2 \cdot 10^{-3} & 0.25 < (\dfrac{r}{10^{17}\, cm}) < 1.24 \\
7.5 \cdot 10^{-4} & 1.24 < (\dfrac{r}{10^{17}\, cm}) < 2.20 \\
5.2 \cdot 10^{-3} & 2.20 < (\dfrac{r}{10^{17}\, cm}) < 5.20 \\
\end{array}
\right\}
\end{equation}

\noindent The corresponding H$_2$ gas density as a function of radius is plotted in Figure\,7.  For comparison, we also plot in this figure the dependence in H$_2$ gas density with radius as derived by \citet{cc07} and \citet{dvt09} for the CO-emitting gas.  The {\siom} abundance in each of these three zones is:

\begin{equation}
[SiO]/ [H_2] = \left\{
\begin{array}{lr}
9 \cdot 10^{-7} (\dfrac{r}{10^{17}\, cm})& 0.25 < (\dfrac{r}{10^{17} cm}) < 1.24 \\
4.5 \cdot 10^{-6} & 1.24 < (\dfrac{r}{10^{17}\, cm}) < 2.20 \\
4.8 \cdot 10^{-8} (4.8 - \dfrac{r}{10^{17}\, cm}) & 2.20 < (\dfrac{r}{10^{17}\, cm}) < 5.20 \ \\
\end{array}
\right\}
\end{equation}

\noindent as plotted in Figure\,8.  For a constant mass-loss rate as assumed in each of the three zones, the {\siom} abundance is required to increase with increasing radius in the inner zone, be approximately constant in the middle zone, and decrease with radius in the outer zone.  An alternative, more physical, way to interpret the change in {\siom} abundance with radius is a corresponding change in the filling factor of the SiO-emitting gas as a function of radius.  The predicted radial profiles in brightness temperatures of the {\sioi} and {\sioii} lines are shown in Figure\,9, and their corresponding line ratio in Figure\,10.  Figure\,11 compares our model predictions for the integrated spectra in {\sioii}, {\sioiii}, and {\siov} (solid line) compared with measurements (histogram) made with the IRAM 30-m telescope by \citet{ql16}.

\subsection{Caveats}
The {\sioi} and {\sioii} measurements used in our model are separated in time by more than 10 years, raising  the possibility that one or both lines have changed in intensity during the intervening period.  A comparison between single-dish observations of the {\sioii} and {\sioiii} lines in 1993 by Bujarrabal et al. (1994) and those in 2006 by Quintana-Lacaci et al. (2007) indicates a possible change in the intensity of the {\sioiii} line at the level of $\sim$20\%. From line surveys of IRC+10420 at 3\,mm and 1\,mm, Quintana-Lacaci et al. (2016) found that while most of the molecular lines observed seem to be stable over time, relatively high $J$ transitions of SiO and HCN such as {\sioiii} and HCN\,(J = 3-2) show noticeable variations over time. Recently, de Vicente et al. (2016) reported that the {\sioi} emission from IRC+10420 is slightly variable in intensity, changing by roughly 20\% over an interval of 1 year.  We note, however, that the uncertainty in flux calibration for single-dish telescopes is quite large, about 10\% at the lower frequency of the {\sioi} line and about 30\% at the higher frequency of the {\siov} line \citep{ql16}.  Systematic uncertainties in the flux calibration, a problem that also plagues our observation in {\sioi} as described in $\S\ref{flux calibration}$, can dominate over any intrinsic changes in line intensity.

By fitting the spectral energy distribution of IRC+10420 measured with the Infrared Space Observatory (ISO), Quintana-Lacaci et al. (2016) inferred an equivalent black body temperature of 300\,K for the dust shell around this star.  The derived temperature is significantly lower than that used in our model of 400\,K, as described in $\S\ref{model assumptions}$; we note that the higher temperature was inferred by fitting the spectral energy distribution of IRC+10420 over a much broader range in wavelengths from infrared to millimeter.  To estimate the sensitivity of our model to changes in the infrared radiation field inside the envelope, we have re-run our calculations for a lower equivalent black body temperature of 300\,K for the dust shell, but keeping all the other physical parameters for the SiO-emitting envelope the same as described in $\S\ref{model results}$.  In Figure\,12, we show our model predictions for the integrated intensity profiles of {\sioii}, {\sioiii}, and {\siov}.  As can be seen, the lower transitions of {\siom} are much less affected by changes in the infrared radiation field than the higher transitions, as radiative excitation contributes more strongly to the overall excitation of {\siom} at higher than at lower transitions.  Specifically, the peak intensity of {\sioii} weakens by only $\sim$10\%, whereas that of {\siov} weakens by $\sim$30\%.  Given the additional uncertainties associated with flux calibration in both interferometric and single-dish observations, we regard our model to provide an acceptable match with currently available observations in all {\siom} lines.

Finally, bear in mind the simplifying assumption made that the SiO-emitting envelope is spherically symmetry, despite the clear departure from circular symmetry and the clumpiness seen in our higher angular resolution map in {\sioi}.  Our model only provides the azimuthally averaged radial density profile and SiO abundance of the envelope.  In regions where the brightness temperature in either {\sioi} or {\sioii}, or both, differ from the azimuthally-averaged value, the physical conditions might well be different from that indicated by our model.  

\section{Discussion}\label{discussion}

\subsection{Dense SiO-emitting clumps within diffuse CO-emitting gas}
Our modeling results clearly show that the SiO-emitting gas has a much higher molecular (primarily H$_2$) gas density than that derived by either \citet{cc07} and \citet{dvt09} for the CO-emitting gas throughout much of the envelope.  The most striking difference is found for the middle zone, which corresponds to a gap devoid of CO-emitting gas between the inner and outer CO-emitting shells in the models of both \citet{cc07} and \citet{dvt09}.  In this zone, the inferred density of the SiO-emitting gas is $\sim$10$^{4}$ cm$^{-3}$ (Fig.\,8), only somewhat lower than that in the inner and outer zones.  In the outer zone, the density of the SiO-emitting gas is about an order of magnitude higher than that obtained by \citet{dvt09} and about two orders of magnitude higher than that obtained by \citet{cc07} for the CO-emitting gas.  In the inner zone, corresponding to the inner CO shell, the density of the SiO-emitting gas is comparable to or only somewhat elevated compared to that of the CO-emitting gas.

Despite the obvious differences in gas density responsible for {\siom} and CO emission throughout much of the envelope of IRC+10420, the SiO-emitting gas spans essentially the same radius from the star as the CO-emitting gas.  We therefore conclude that the {\siom} emission, particularly that from the middle and outer zones, must originate from relatively dense clumps.  By comparison, the relatively diffuse gas between these clumps, which contribute little to the observed {\siom} emission, gives rise predominantly to the observed CO emission.  At this point, we note that the gap between the inner and outer CO-emitting shells inferred by \citet{cc07} and \citet{dvt09} may be a consequence of an oversimplification of their models for reproducing the overall CO-emitting envelope (i.e., a minimum parameter model).  The observed CO emission shows no such gap, but instead just a weak local depression at the angular radius of the purported gap \citep[see Fig.\ 10 of][]{cc07} or \citep[Figs.\, 10 and 14 of][]{dvt09}.  Decreasing the stellar mass-loss rate for producing the inner and/or outer CO-emitting shells in the models of \citet{cc07} and \citet{dvt09} would necessitate increasing the stellar mass-loss rate during the interval between the production of these two shells.  In short, the CO observations indicate a drop, but not necessarily a null, in the stellar mass-loss rate between the production of the inner and outer shells.  By contrast with the middle and outer zones, in the inner zone where the density of the CO-emitting gas is relatively high and comparable to the critical density of {\sioi} if not {\sioii}, the same gas component may be responsible for both the observed CO and SiO emission.  In Figure\,13, we sketch a qualitative model for the origin of the CO and {\siom}  emission from the envelope of IRC+10420.

Which component has the higher filling factor, the dense primarily SiO-emitting or diffuse primarily CO-emitting gas?   At a given radius, the azimuthally averaged brightness temperatures in {\sioi} and {\sioii} are roughly comparable (within a factor of $\sim$2) with those in CO\,(J = 1-0) and CO\,(J = 2-1) \cite[see Figs.\,13--14 of][]{dvt09}.  Their roughly comparable brightness temperatures suggest roughly comparable filling factors for the two components, assuming that both components have the same temperature (as we have in our model for the SiO-emitting envelope).  If the dense clumps also have higher temperatures, a possibility we explore below to explain the presence of SiO in the gas phase so far out in the envelope of IRC+10420 ($\S\ref{origin SiO}$), then their filling factor decreases.  A more detailed estimate of the filling factor of the two gas components would require modelling of the CO emission from both the dense clumps and diffuse interclump gas.

As can be seen in Figure\,7, the SiO abundance is highest in the middle zone.  By comparison, the SiO abundance is dramatically lower in the inner and outer zones, with the outer zone having the lowest abundance.  The modelled SiO abundance implicitly assumes a unity filling factor for the SiO-emitting gas, and therefore represents a lower limit in the intrinsic SiO abundance (i.e., the intrinsic SiO abundance scales in proportion to the inverse of the filling factor).  If the intrinsic SiO abundance in the three zones are similar, then the different SiO abundances inferred in our model imply different filling factor for the dense SiO-emitting clumps in these zones.  As mentioned above, in the inner zone, the same gas is likely responsible for both the SiO and CO emission.  If the gas filling factor is unity in this region, then the SiO abundance inferred in our model for the inner zone reflects the intrinsic SiO abundance in this zone.  In that case, the intrinsic SiO abundance must be much higher in the middle compared with the inner zone.

\subsection{Shock origin of SiO}\label{origin SiO}
According to current chemical models, SiO molecules form in stellar atmospheres under local thermal equilibrium (Cherchneff 2006).   These (and other) molecules are then ejected through collective collisional coupling with dust grains driven outwards by radiation pressure from the star.  As the temperature drops with radial distance, SiO molecules are expected to condense onto dust grains, leading to an abrupt decrease in SiO abundance beyond a certain radius.  In the model proposed by Jura \& Morris (1983), the SiO abundance depends on the competition between sticking onto and evaporation from the surface of dust grains.  The SiO evaporation rate has an approximately exponential dependence on the grain temperature, so that at low temperatures and high gas densities, SiO molecules mainly stick to dust grains. 

Jura \& Morris (1983) found that the condensation of molecules becomes dominant
when the grain temperature drops below T$_{bind}$/50, where kT$_{bind}$ is the binding kinetic energy of molecules on the grains.  For SiO molecules, T$_{bind} = 29500$\,K, implying that SiO condenses onto dust grains when the grain temperature drops below $\sim$600\,K.  In AGB stars, temperatures above the SiO condensation temperature are usually found only at the inner region of their envelopes.  As a consequence, SiO emission is confined to a relatively small radial distance (by comparison with CO emission) from these stars.  Further out, the SiO abundance drops dramatically as condensation becomes dominant and almost every molecule that sticks onto the grain remains there.  Beyond the radius where evaporation becomes negligible, the fractional abundance of SiO molecules as a function of radius, $f_{\rm SiO}(r)$, can be determined from the equation given in Jura \& Morris (1983):

\begin{equation} 
f_{\rm SiO}(r) = f_{\rm SiO}(r_{\rm o}) {\rm \ exp}[-\delta (\frac{1}{r_{\rm o}} - \frac{1}{\rm r})]
\end{equation}
where $f_{\rm SiO}(r_o)$ is the fractional abundance at $r_{\rm o}$, the condensation radius, and $\delta$ is the characteristic scale length.  The latter is defined as:

\begin{equation}
\delta = \frac{\alpha \dot{N_d}\sigma_{\rm gr}v_{dr}}{4\pi v_{\rm e}^2}
\end{equation}
where $\alpha$ is the sticking probability of SiO onto dust grains, $\dot{N_d}$ is the dust mass-loss rate in terms of dust grain number, $\sigma_{\rm gr}$ is the grain cross section, and $v_{\rm dr}$ is the drift velocity of the dust grains with respect to the gas.  The condensation radius ($r_{\rm o}$) is determined from the manner in which the temperature of the dust grains decreases with increasing radius.  This simple model was used by Gonzalez Delgado et al. (2003) to successfully explain the observational characteristics of SiO emission from both carbon-rich and oxygen-rich circumstellar envelopes.  

Using the parameters of the CO-emitting envelope as derived by \citet{dvt09}, we estimate a condensation radius of $r_{\rm o} \sim 4 \times 10^{15}$\,cm ($\sim$300\,AU) for IRC+10420, significantly smaller than the inner radius of the inner zone in our model.  The estimated condensation radius is about an order of magnitude larger than that for AGB stars, as a consequence of the higher temperature in the envelope of IRC+10420 (given its much higher luminosity) at a given radius from the central star.  Assuming a sticking probability of $\alpha \sim 1$ (as is adopted for, and which explains the radial extent of the SiO emission around, AGB stars), the characteristic scale length is therefore $\sim$$1.8 \times 10^{17}$\,cm, which is much larger than the condensation radius (note that the larger scale length, the more quickly the SiO abundance decreases beyond the condensation radius).  
Thus, the SiO abundance is predicted to drop precipitously beyond the condensation radius, in stark contrast with the dramatic outward increase in the SiO abundance inferred from our model between the inner and middle zones.  Indeed, the SiO abundance is predicted to become vanishingly small within the inner zone of the SiO-emitting envelope, let alone the middle and outer zones.

The much higher SiO abundance throughout the SiO-emitting envelope of IRC+10420 than expected due to condensation onto dust grains makes necessary a mechanism for evaporating SiO molecules from the surfaces of these grains.  The most obvious such mechanism is sputtering (collisions between gas and dust due to their slightly different outflow velocites) or shocks, as is conjectured for SiO emission detected from protostellar outflows as well as fast outflows from post-AGB stars (where, once again, any SiO in the gas phase is expected to condense onto dust grains near the star).  Shocks may also explain the relatively high turbulent velocity in the envelope of IRC+10420, as noted by \citet{nedoluha92} based on the broad and irregular line shapes of OH masers observed from this star.
We note that the sound speed in the molecular envelope is only ~1 kms$^{-1}$, and hence a difference in velocities of 
just a few kms$^{-1}$ or higher between the SiO-emitting clumps and surrounding CO-emitting gas is sufficient to generate strong shocks.

\subsection{Relationship dense SiO-emitting to optical dust clumps}
The inference of dense gas clumps responsible for the {\sioi} and {\sioii} emission of IRC+10420 naturally suggests a connection with dust clumps seen in optical scattered light around this star (Humphreys et al. 1997 and Tiffany et al. 2010).  Indeed, such dust clumps are detectable out to an angular radius of $\sim$2\arcsec\ from the star, roughly comparable in angular extent to the middle zone at or near where the SiO emission is brightest. Humphreys et al. (1997) and Tiffany et al. (2010) suggest that these clumps are ejected as a consequence of violent global events on the surface of the star, probably related to convective or magnetic activities.  We speculate that as these dense clumps move outwards, shocks can form due to interaction between these clumps and the surrounding diffuse CO-emitting gas where the difference in their bulk velocities is as low as just a few kms$^{-1}$.  Heating due to such shocks leads to an increase in the temperature of the grains, releasing SiO molecules from the surfaces of these grains back into the gas phase. 

From dual-epoch observations with the HST, Tiffany et al. (2010) determine the proper motion, and thus the transverse
velocity, for a number of dust clumps in the envelope of IRC+10420. By combining the transverse velocity and the light of sight velocity (Humphreys et al. 2002), they find a space motion for these clumps 
close to the plane of the sky. As a consequence, Tiffany et al. (2010) suggest that the ejecta is viewed nearly pole-on.
Indeed, imaging polarimetry by Shenoy et al. (2015) reveal a concentric pattern in the linear polarization of the scattered light, with fractional linear polarization of up to 30\% at distance of 1\arcsec--2\arcsec\ from the star.  Such high degrees of linear polarization requires a scattering angle close to 90\degr; i.e., the bulk of the scatterers are distributed near the sky plane.
In addition, optical interferometric observation of Br$\gamma$ emission by \citet{oudmaijer13} is found to be consistent with a bipolar 
outflow confined within $\sim$10 AU around the central hypergiant star and viewed nearly pole-on. 
Therefore, all previous observations in the optical and near infrared seem to suggest a strong departure from spherical symmetry of the
envelope. By contrast, the dense SiO-emitting clumps inferred from our observations and analysis are distributed throughout a roughly 
spherical envelope. The reason for such a discrepancy between optical and radio observations currently remains unresolved
and clarification will require further investigation.

\section{Summary and Conclusion}\label{summary}

We presented an image, taken with the VLA, of the circumstellar envelope of the yellow hypergiant IRC+10420 as traced in {\sioi}.  Like that traced in {\sioii} as reported by \citet{cc01}, the global kinematic structure of the envelope traced in the SiO lines is similar to that traced in both CO(1-0) by \citet{dvt09} and CO(2-1) by \citet{cc01}, including a weak velocity gradient along the north-west to south-east direction apparent in all the aforementioned molecular lines.  By contrast with the envelope traced in CO, however, both the {\sioi} and {\sioii} emissions show a pronounced central depression that giving rise to their ring-like appearance.  The ring traced in {\sioi} is clumpy and brighter on the north-western than south-eastern side, along the same direction as the aforementioned velocity gradient in the envelope.  Other than that, on a global scale, the envelope is approximately spherically symmetric as traced in both {\sioi} and {\sioii}, as is the case as traced in both {\com} $J = 1-0$ and {\com} $J = 2-1$.  A detailed comparison between the {\sioi} and {\sioii} images reveal that:

\begin{itemize}

\item[1.]  the {\sioi} emission peaks at a radius of $\sim$2\arcsec, whereas the {\sioii} emission peaks at a significantly smaller radius of $\sim$1\arcsec.

\item[2.]  the ratio in brightness temperature between {\sioii} and {\sioi}, {\sioii}/{\sioi}, is significantly higher within a radius of $\sim$2\arcsec, rising to a value as high as about 2.2 at the innermost region, than further out, where the line ratio is approximately constant at about 1.0.  

\end{itemize}

\noindent Although systematic errors in the flux calibration introduce unknown uncertainties in the measured brightness temperatures of both the SiO lines, such errors do not change the observed radial trends in the brightness temperatures or the ratio in brightness temperatures of both these lines.  While such errors may lead to a quantitative change in the physical parameters we derive for the SiO-emitting envelope, they do not change our conclusions on the qualitative differences in physical properties between the SiO-emitting and CO-emitting parts of the envelope.

We constructed a model that reproduces both the measured radial dependence in {\sioii}/{\sioi} and the brightness temperatures in both {\sioi} and {\sioii}. Because these measurements do not provide meaningful constraints on the gas temperature, with {\sioii}/{\sioi} in particular being relatively insensitive to temperature, we assumed a temperature profile for the envelope as inferred by \citet{teyssier12} from CO observations (for the inner region of the envelope), or as adopted by \citet{cc07} for modelling of the CO emission (for the outer region of the envelope), in the manner parameterised by Eq.\,3.  In our model, we divided the SiO-emitting envelope into three zones as is inferred for the CO-emitting envelope in the models of both \citet{cc07} and \citet{dvt09}.  In, specifically, the model of \citet{cc07}, the inner zone spans the radial range $\sim$0\farcs3--1\farcs5, the middle zone $\sim$1\farcs5--2\farcs5, and the outer zone $\sim$1\farcs5--6\farcs0.  The middle zone is essentially devoid of CO-emitting gas in the models of both \citet{cc07} and \citet{dvt09}, indicating a dramatic drop in H$_2$-gas density or its filling factor in this zone.  We find that:

\begin{itemize}

\item[3.] except at the inner zone, the H$_2$-gas density inferred for the CO-emitting envelope is far too low to produce the observed brightness temperatures in both {\sioi} and {\sioii}.

\item[4.]  in the outer zone, the H$_2$-gas density traced by {\sioi} and {\sioii} is an order of magnitude higher than that inferred by \citet{dvt09} and two orders of magnitude higher than that inferred by \citet{cc07} for the CO-emitting envelope.  

\item[5.]  in the middle zone, where {\sioi} peaks in brightness temperature, the H$_2$-gas density traced by {\sioi} and {\sioii}  is no more than an order of magnitude lower than that of the outer zone, by contrast with the lack of CO-emitting gas in this zone as inferred by both \citet{cc07} and \citet{dvt09}.  

\end{itemize}

We conclude that the {\siom} emission, particularly that from the middle and outer zones, must originate from relatively dense clumps.  By comparison, the relatively diffuse gas between these clumps, which contribute little to the observed {\siom} emission, gives rise predominantly to the observed CO emission.  If both gas components are at the same temperature, as in our model, the filling factors of the two gas components must be roughly comparable; a more detailed estimate of the filling factor of the two gas components would require modelling of the CO emission from both the dense clumps and diffuse interclump gas.  On the other hand, a higher temperature for the dense {\siom}-emitting clumps imply a lower filling factor for these clumps.  Our model requires the {\siom} abundance to change with radius in each zone.  Interpreting this change as an actual change in the filling factor of the SiO-emitting gas having a constant intrinsic {\siom} abundance throughout the envelope:

\begin{itemize}

\item[6.]  the filling factor of the SiO-emitting gas increases with radius in the inner zone, is nearly an order of magnitude higher and remains approximately constant with radius in the middle zone, and is lowest and moreover decreases with increasing radius in the outer zone

\end{itemize}

Despite the relatively high effective temperature and luminosity of IRC+10420, we show that SiO molecules are expected to condense onto dust grains just beyond the inner radius of the inner zone, and to become vanishingly small within this zone.  We speculate that SiO molecules are subsequently released from dust grains due to shock interactions between the dense SiO-emitting clumps and the diffuse CO-emitting interclump gas, where differences in bulk velocity as small as a few kms$^{-1}$ is sufficient to generate strong shocks. We associate the dense SiO-emitting clumps with dust clumps seen in scattered optical light.  Such dust clumps are detectable out to an angular radius of $\sim$2\arcsec\ from the star, roughly comparable in angular extent to the middle zone at or near where the SiO emission is brightest.

We thank Dr. A. Castro-Carrizo for providing the PdBI {\sioii} data. Dinh-V-Trung acknowledges the 
financial support of the National Foundation for Science and Technology Development (Nafosted) under grant number 103.99-2014.82.  
K.~T.~Wong thanks The University of Hong Kong (HKU) for the support of
a postgraduate studentship. Part of the work was submitted to HKU by
K.~T.~Wong for the award of the degree of Master of Philosophy.
This research has made use of NASA's Astrophysics Data 
System Bibliographic Services and the SIMBAD database operated at CDS, Strasbourg, France.

{\it Facilities:} \facility{VLA}.



\clearpage

\begin{table}[ht]
\centering
\caption{VLA D-array observations of quasar 3C\,345 in Q-band around 19 April 2010.\label{table1}}
\begin{tabular}{|c|c|c|p{110pt}|p{110pt}|}
\hline
Observation & Date & Project & Scientific Object(s) & Q-band Flux (7mm) of 3C\,345 (VLA) \\ \hline
1 & 10 April 2010 & S2053 & quasars & $6.0672 \pm 0.0004$ Jy \\ \hline
2 & 19 April 2010 & AD621 & IRC+10420 & 7.38 Jy (interpolated) \\ \hline
3 & 22 April 2010 & AD621 & AFGL2343 & --- \\ \hline
4 & 26 April 2010 & AD621 & AFGL2343 & $8.355 \pm  0.001$ Jy \\ \hline
\end{tabular}
\end{table}
\begin{table}[ht]
\centering
\caption{Total Q-band (7mm) flux of 3C\,345 from VLBA Blazar Monitoring Project.}
\label{table2}
\begin{tabular}{|c|c|}
\hline
Date & Q-band Flux (7mm) of 3C\,345 (VLBA) \\ \hline
6 March 2010 & $5.470 \pm 0.005$ Jy \\ \hline
6 April 2010 & $6.180 \pm 0.005$ Jy \\ \hline
14 April 2010 & $5.836 \pm 0.004$ Jy \\ \hline
19 May 2010 & $5.049 \pm 0.004$ Jy \\ \hline
14 June 2010 & $5.077 \pm 0.004$ Jy \\ \hline
\end{tabular}
\end{table} 
\newpage
\begin{figure}
  \centering
  \includegraphics[width=\textwidth]{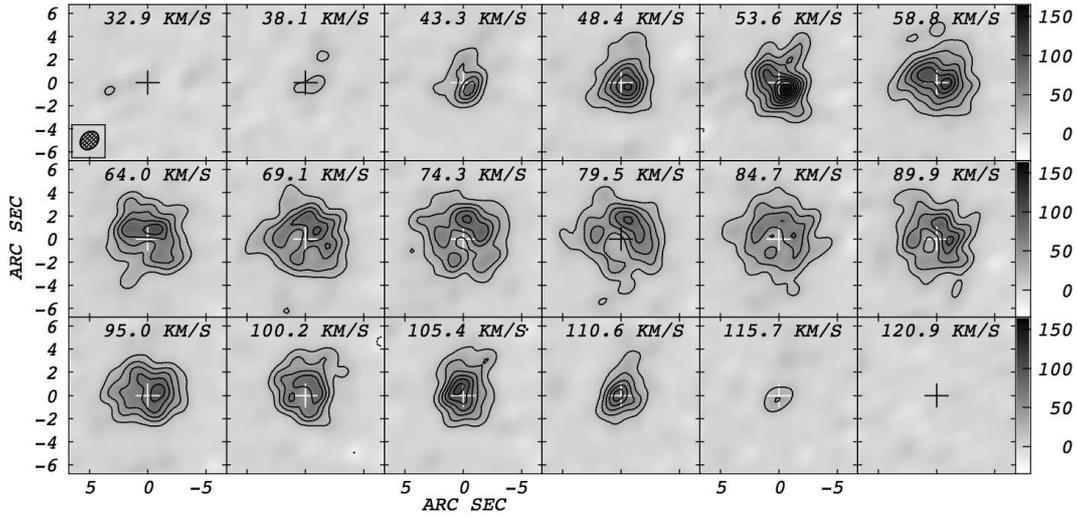}
  \caption[Channel maps of {\sioi} emission]{VLA maps of {\sioi} thermal emission from IRC+10420. 
  LSR Velocities are indicated in each channel map. Each contour level is $3 \sigma = 17.7$ mJy. 
  The greyscale wedge represents the SiO intensity in mJy. The FWHM of the synthesized beam 
  ($1''.78 \times 1''.36$, $\text{PA} = -44.51\dg$) is drawn at the top-left corner of the first 
  channel map. North is up and east is left. The cross represents the position of IRC+10420, 
  estimated from 1.3 mm continuum emission {\protect\citep{dvt09}}.}
  \label{fig1}
\end{figure}

\newpage

\begin{figure}
  \centering
  \includegraphics[width=\textwidth]{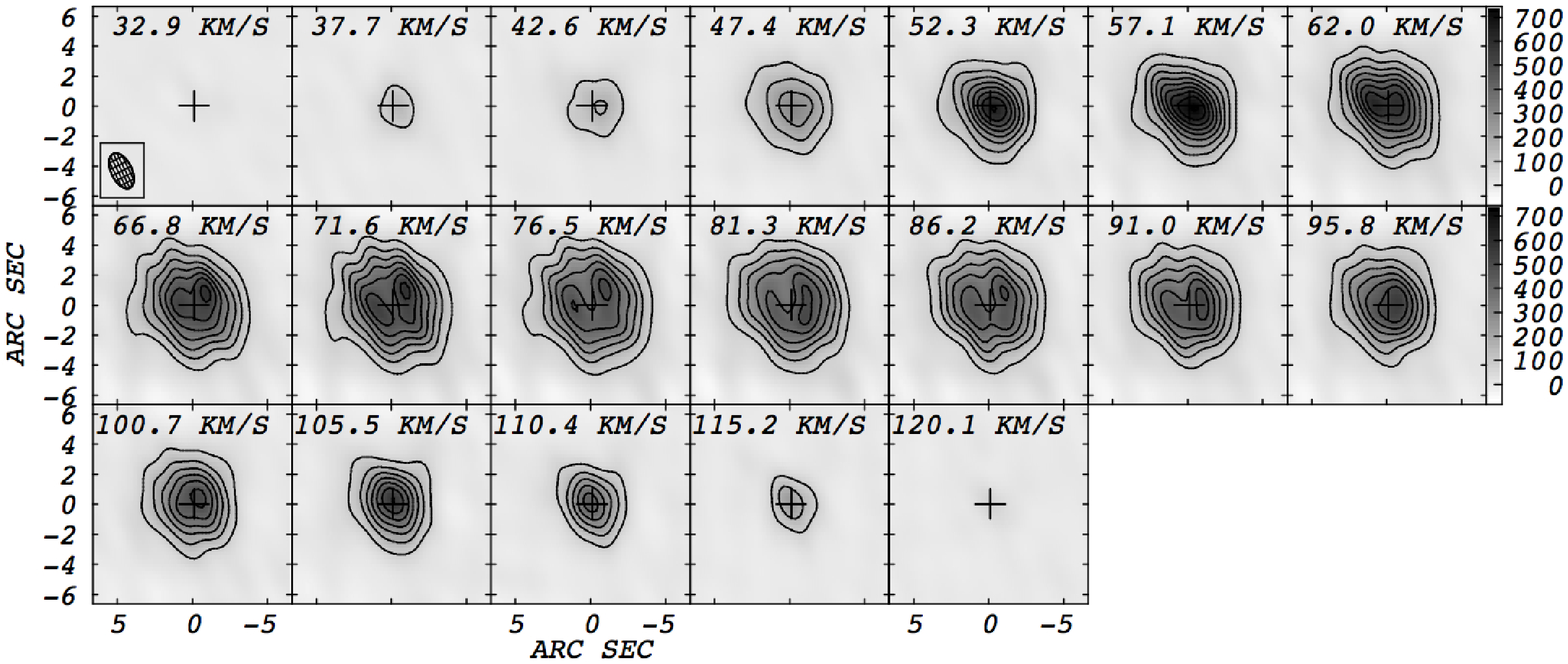}
  \caption[Channel maps of {\sioii} emission]{Regridded PdBI maps of {\sioii} thermal emission from IRC+10420, 
  originally published in {\protect\citet{cc01}} into the same channel spacing as in 
  Figure \ref{fig1} (${\approx}4.84 \kms$). LSR Velocities are indicated in each channel map. 
  Each contour level is 80 mJy. The greyscale wedge represents the SiO intensity in mJy. 
  The FWHM of the synthesized beam ($2''.53 \times 1''.38$, $\text{PA} = 25.98\dg$) is drawn 
  at the top-left corner of the first channel map. North is up and east is left. The cross 
  represents the position of IRC+10420, estimated from 1.3 mm continuum emission {\protect\citep{dvt09}}.}
  \label{fig2}
\end{figure}

\newpage

\begin{figure}
  \centering  
  \includegraphics[width=\textwidth]{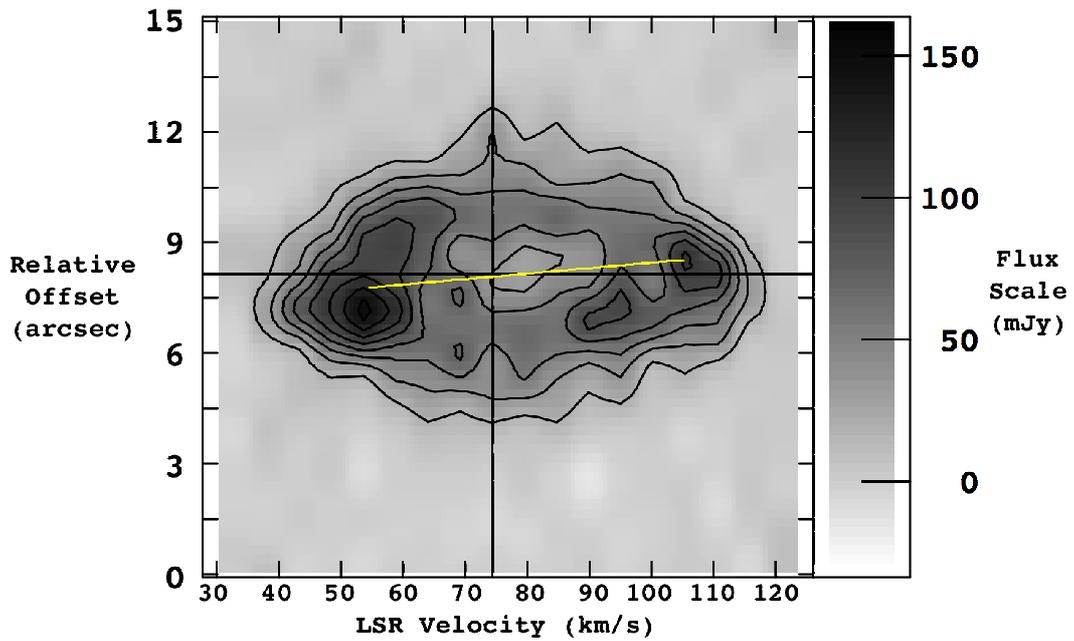}
  \caption[Position-velocity (PV) diagram of {\sioi} emission]{Position-velocity (PV) diagram of 
  the VLA observation on {\protect\sioi} along PA $=70\dg$. Horizontal axis represents the LSR 
  Velocities in km s$^{-1}$; vertical axis represents the relative offset, from south-west ($0''$) 
  to north-east ($15''$) along the $70\dg$-axis. Greyscale represents the intensity of emission in 
  the unit of milli-Jansky. Position and systemic velocity of IRC+10420 
  (RA $=19^{\text{h}}26^{\text{m}}48^{\text{s}}.09$, $V_{\text{LSR}}=73.9 kms^{-1}$) 
  are indicated by the horizontal and vertical reference lines respectively. The velocity 
  gradient of {\com} envelope observed by {\protect\citet{cc07}} is indicated by the yellow line.}
  \label{fig3}
\end{figure}

\newpage

\begin{figure}
  \centering
  \includegraphics[width=\textwidth]{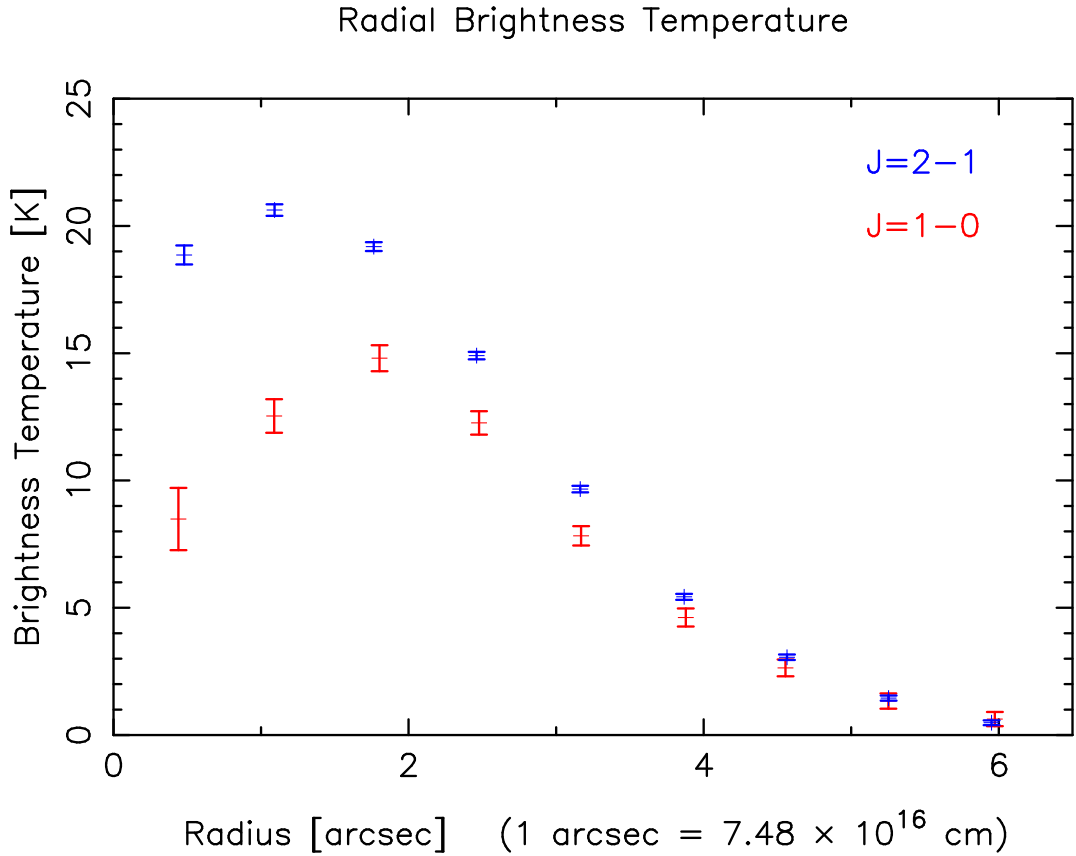}
  \caption[Radial brightness temperature profiles]{Azimuthally averaged radial brightness 
  temperature profiles for {\sioi} and {\sioii} . Conversion factor and map noise of 
  {\sioi} are 185.63 K/Jy, $1.1 \text{ K/beam} = 5.9$ mJy/beam, and those of {\sioii} 
  are 46.47 K/Jy, $0.26 \text{ K/beam} = 5.5$ mJy/beam respectively. Uncertainty bars 
  have been scaled as the inverse square root of the annular ring coverage (in unit of synthesized beams).}
  \label{fig4}
\end{figure}

\newpage

\begin{figure}
  \centering
  \includegraphics[width=\textwidth]{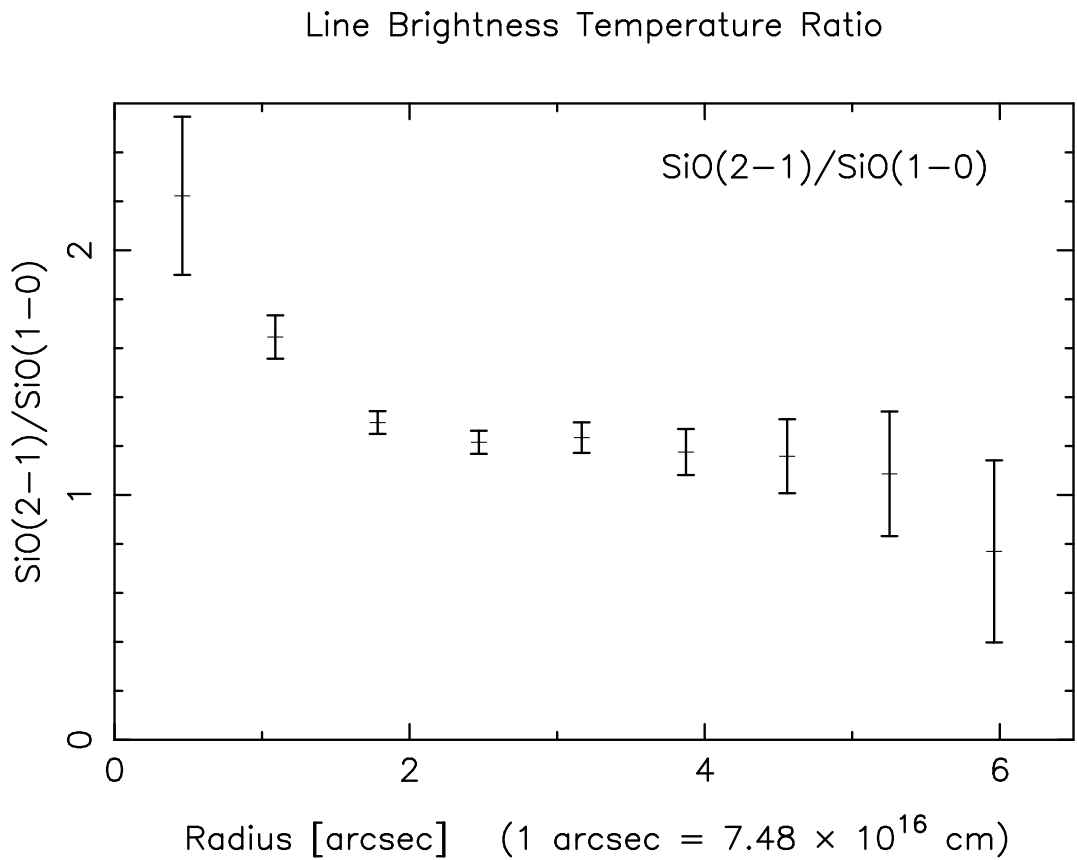}
  \caption[Radial line ratio profile]{Line brightness temperature ratio of {\sioii} over {\sioi} 
  computed from the temperature profile as shown in Figure {\protect\ref{fig4}}. 
  Uncertainty bars have been calculated in standard manner of error propagation.}
  \label{fig5}
\end{figure}

\newpage

\begin{figure}
  \centering
    \includegraphics[width=0.85\textwidth]{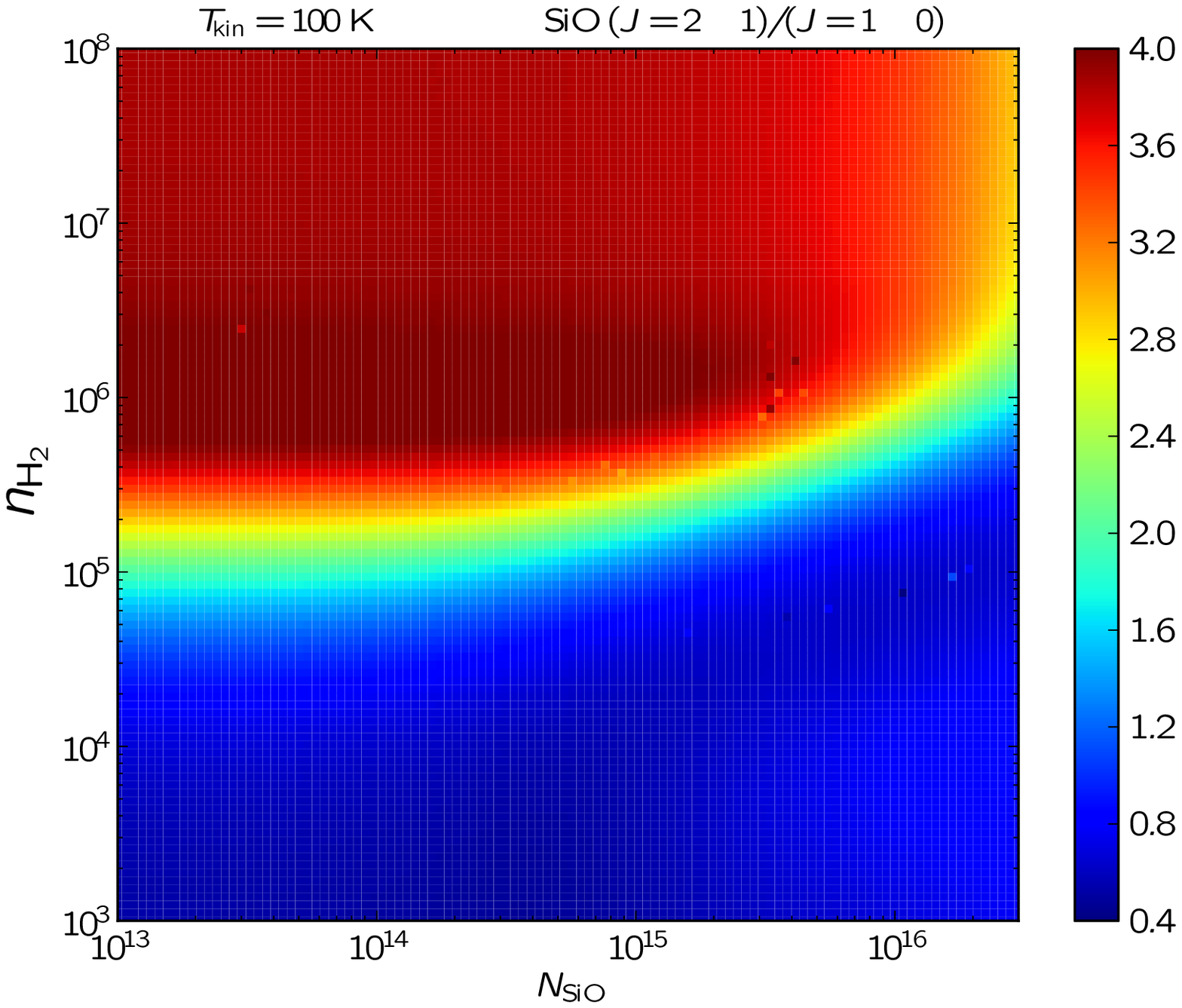}
  \caption[\texttt{RADEX} simulation of line ratio with respect to density and {\siom} 
  column density]{Simulated line brightness temperature ratio (the colour scale) of 
  {\sioii} over {\sioi} as a function of number density of molecular hydrogen gas 
  (vertical axis: $n_{{\text{H}}_2}$, cm$^{-3}$) and \emph{column density of SiO molecules 
  along the line of sight} (horizontal axis: $N_{\text{SiO}}$, cm$^{-2}$), at a gas 
  temperature of 100 K. Results are calculated by non-LTE code 
  \texttt{RADEX} {\protect\citep{radex}} using large velocity gradient (LVG) approximation 
  for escape probability.}
  \label{fig6}
\end{figure}

\newpage

\begin{figure}
  \centering
  \includegraphics[width=0.85\textwidth]{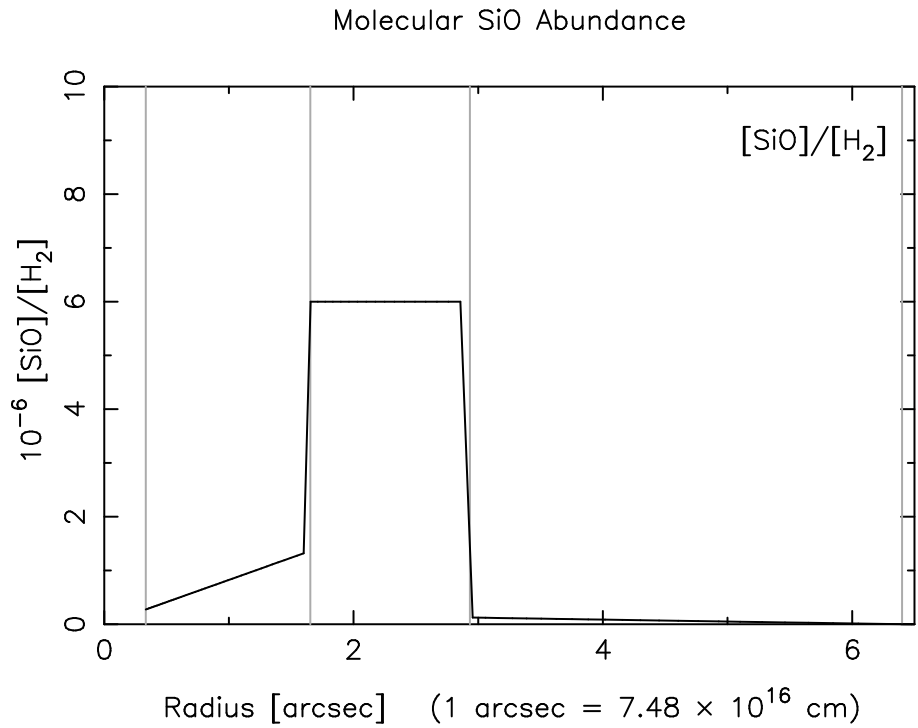}
  \caption[Relative {\siom} molecular abundance in our three-zone model]{Best-fit relative {\siom} 
  molecular abundance profile.  The change in {\siom} abundance with radius may reflect, in part if not entirely, a corresponding change in the filling factor of the SiO-emitting gas as a function of radius.}
  \label{fig:fit-abundance}
\end{figure}

\clearpage

\begin{figure}
  \centering
  \includegraphics[width=0.85\textwidth]{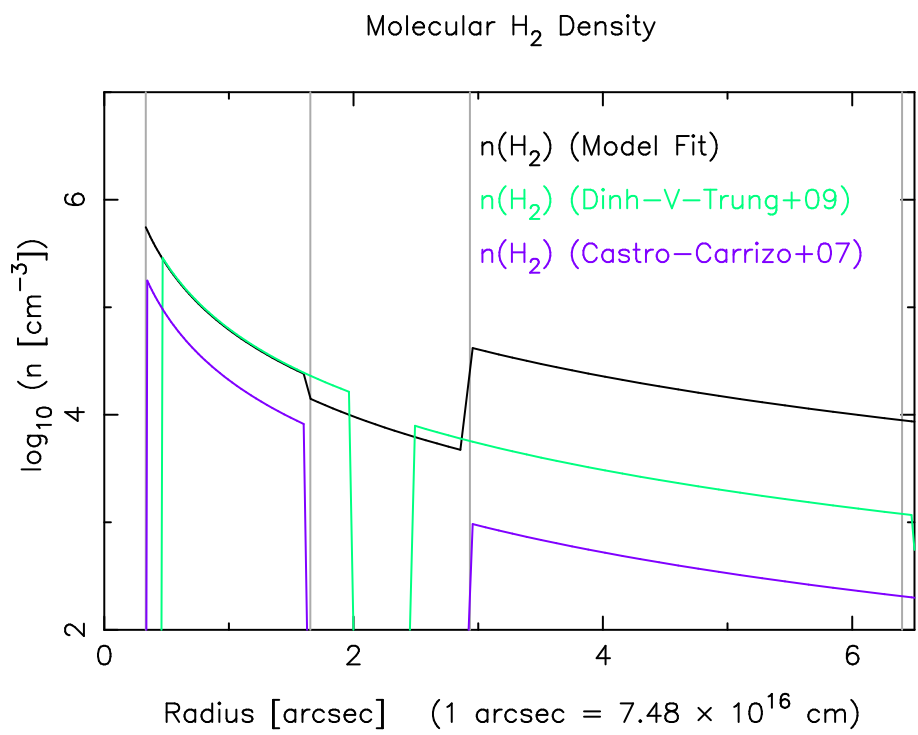}
  \caption[Molecular H$_2$ gas density in our three-zone model]{Best-fit molecular H$_2$ gas density profile.}
  \label{fig:fit-density}
\end{figure}

\clearpage

\begin{figure}
  \centering
  \includegraphics[width=0.85\textwidth]{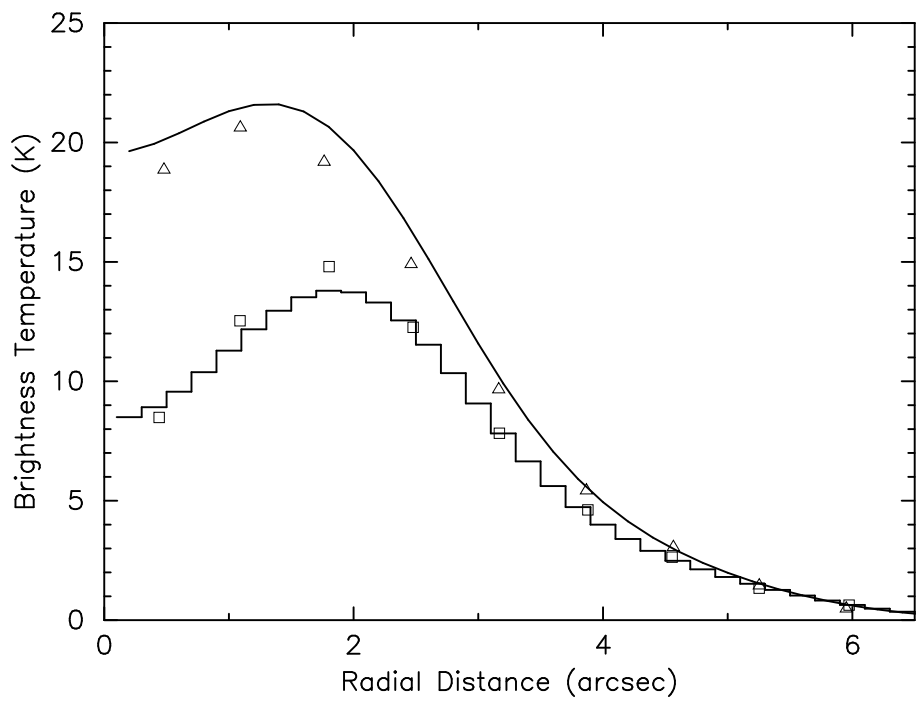}
  \caption[Modelled radial brightness temperature profiles as in our three-zone model]{Model fit of 
  radial brightness temperature profiles under the three-zone model. The points are the observational data .}
  \label{fig:fit-intensity}
\end{figure}

\clearpage

\begin{figure}
  \centering
  \includegraphics[width=0.85\textwidth]{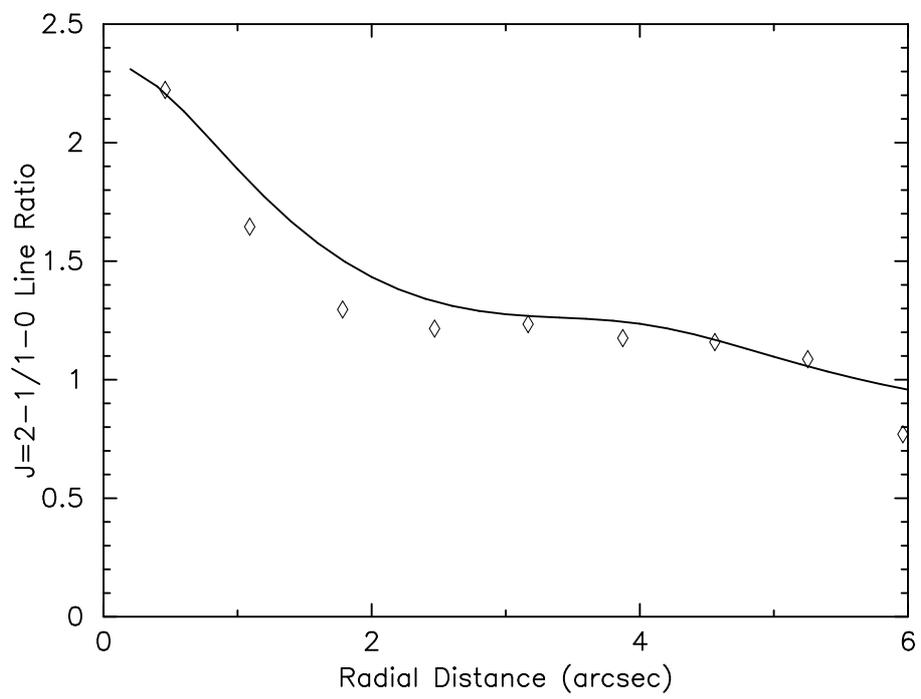}
  \caption[Modelled line ratio as in our three-zone model]{Model fit of radial line brightness 
  temperature ratio profiles under the three-zone model.} 
  \label{fig:fit-ratio}
\end{figure}

\clearpage

\begin{figure}
  \centering
  \includegraphics[width=0.85\textwidth]{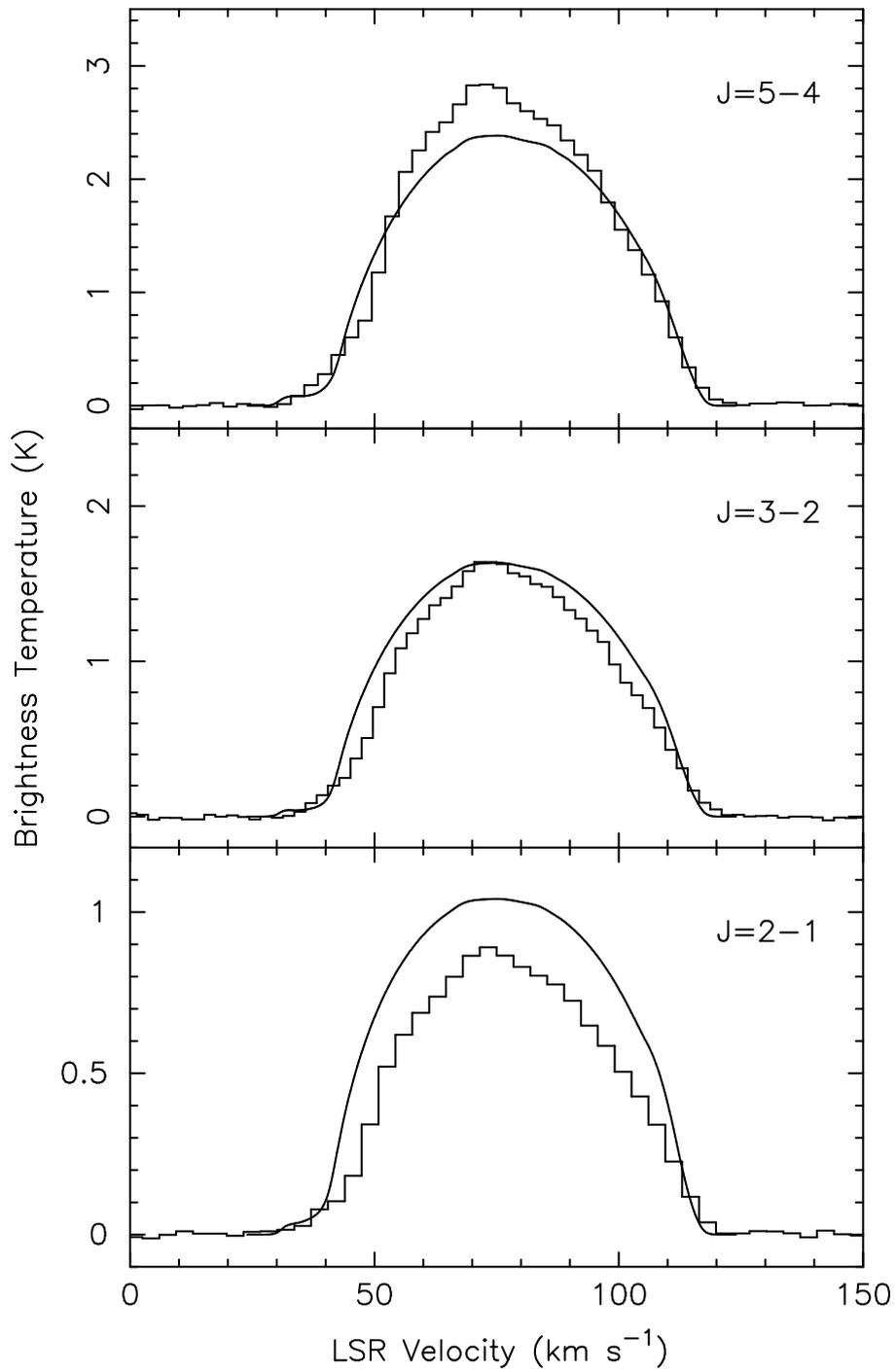}
  \caption[Intensity of SiO rotational lines predicted by our best fit model]{Comparison between modeled SiO rotational lines shown in solid line and the single dish data (Quintana-Lacaci et al. 2007) shown in histogram.}
  \label{fig:singledish}
\end{figure}

\clearpage

\begin{figure}
  \centering
  \includegraphics[width=0.85\textwidth]{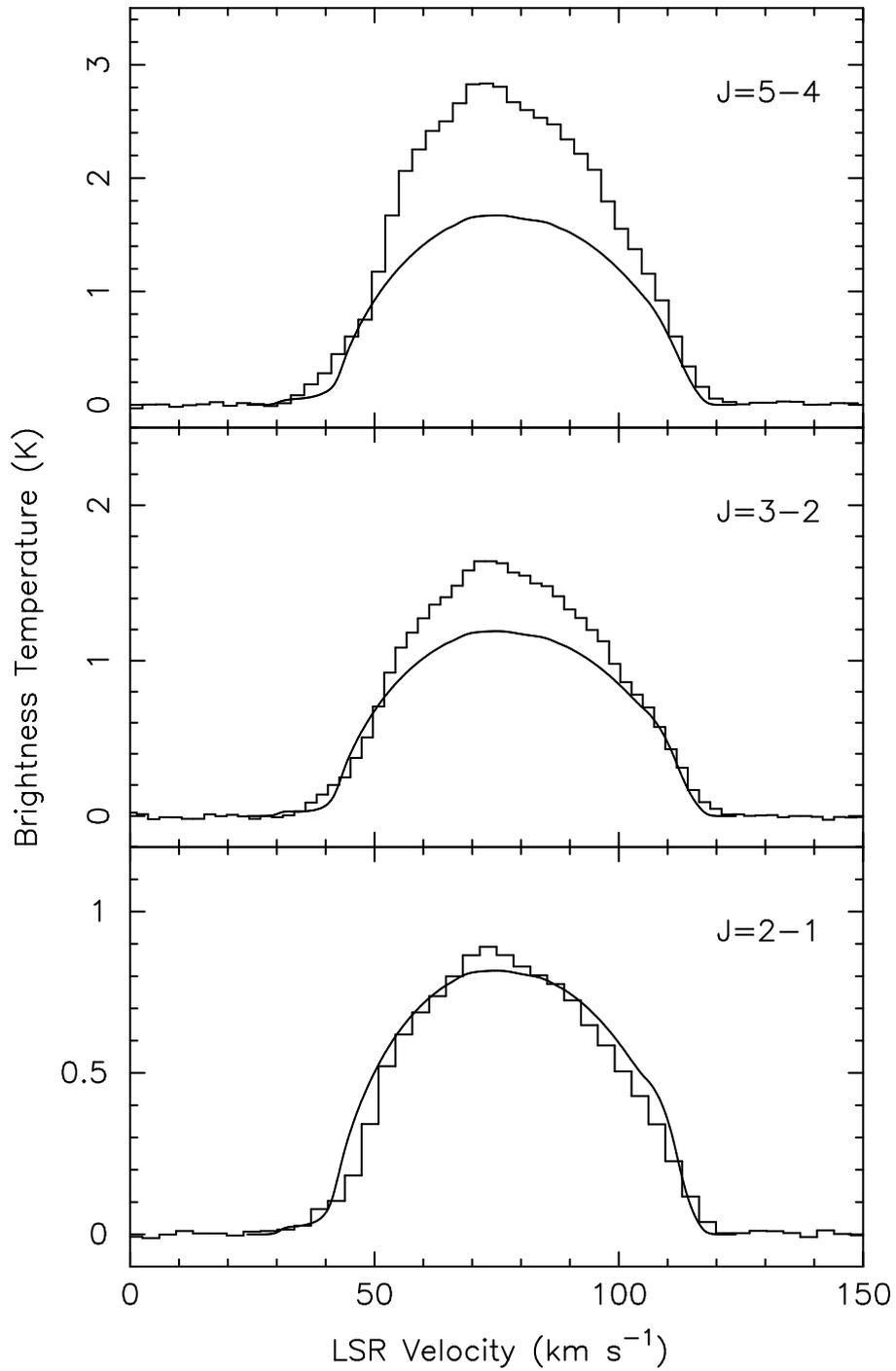}
  \caption[Intensity of SiO rotational lines predicted by our model]{Comparison between modeled SiO rotational lines shown
  in solid line with the 
  temperature of the central radiation source reduced to T=300 K and the single dish data (Quintana-Lacaci et al. 2007) shown in histogram.}
  \label{fig:T300}
\end{figure}

\clearpage

\begin{figure}
  \centering
  \includegraphics[width=0.85\textwidth]{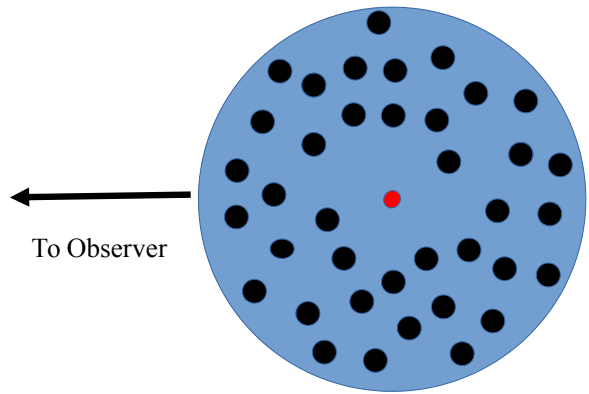}
  \caption[Sketch of envelope]{A schematic view of the structure of the envelope around IRC+10420. Dense SiO emitting clumps are shown as filled black circle. CO emission originates from more diffuse surrounding gas.}
  \label{fig:sketch}
\end{figure}

\end{document}